    \documentclass[prd,aps,preprint,tightenlines,showpacs,nofootinbib,superscriptaddress]{revtex4-1}
\usepackage{mathrsfs}
\usepackage{amsfonts}
\usepackage{amsmath}
\usepackage{amssymb}
\usepackage{array}
\usepackage{verbatim}
\usepackage{bm}
\usepackage{epsfig}
\usepackage{graphicx,color}
\usepackage{relsize}
\usepackage{lineno}
\usepackage{float}
\usepackage{multirow}
\RequirePackage{xspace}

\def\lsim{\raise0.3ex\hbox{$<$\kern-0.75em\raise-1.1ex\hbox{$\sim$}}}

\def\gsim{\raise0.3ex\hbox{$>$\kern-0.75em\raise-1.1ex\hbox{$\sim$}}}

\newcommand{\be}{\begin{equation}}

\newcommand{\ee}{\end{equation}}

\def\beq{\begin{equation}}

\def\eeq{\end{equation}}

\def\beqa{\begin{eqnarray}}

\def\eeqa{\end{eqnarray}}

\newcommand{\ba}{\begin{eqnarray}}

\newcommand{\ea}{\end{eqnarray}}

\def\gappeq{\mathrel{\rlap {\raise.5ex\hbox{$>$}}

{\lower.5ex\hbox{$\sim$}}}}

\def\lappeq{\mathrel{\rlap{\raise.5ex\hbox{$<$}}

{\lower.5ex\hbox{$\sim$}}}}

\def\Toprel#1\over#2{\mathrel{\mathop{#2}\limits^{#1}}}

\begin{document}

\title{Light-by-Light scattering in ultraperipheral heavy ion collisions: Estimating inelastic contributions}

\author{Mariola {\sc Klusek-Gawenda}}
\email{mariola.klusek@ifj.edu.pl}
\affiliation{Institute of Nuclear
Physics Polish Academy of Sciences, ul. Radzikowskiego 152, PL-31-342 
Krak{\'o}w, Poland}

\author{Victor P. {\sc Gon\c{c}alves}}
\email{barros@ufpel.edu.br}
\affiliation{Institute of Physics and Mathematics, Federal University of Pelotas, \\
  Postal Code 354,  96010-900, Pelotas, RS, Brazil}

\author{Antoni {\sc Szczurek}}
\email{antoni.szczurek@ifj.edu.pl}
\affiliation{Institute of Nuclear
Physics Polish Academy of Sciences, ul. Radzikowskiego 152, PL-31-342 
Krak{\'o}w, Poland}
\affiliation{Institute of Physics, Faculty of Exact and Technical Sciences, University of Rzesz\'ow, Pigonia 1 St., 35-310 Rzesz\'ow, Poland\vspace{5mm}}

\begin{abstract}

The current state-of-the-art theoretical estimations lead to cross-sections 
for $AA \to \gamma \gamma AA$ which are somewhat
smaller than the measured ones by the ATLAS and CMS Collaborations,
which motivates the searching and calculation of subleading corrections
disregarded in these previous studies. In this paper, we estimate 
for the first time the contribution of inelastic channels to the Light-by-Light (LbL) scattering in ultraperipheral collisions of heavy ions (UPHICs), in which one or both of the incident nuclei dissociate  ($A A \to \gamma \gamma X Y$ where $X, Y = A, A'$) due to the photon emission. These new mechanisms are related to extra emissions that are rather 
difficult to identify at the LHC and  may be misinterpreted 
as enhanced $\gamma \gamma \to \gamma \gamma$ scattering
compared to the Standard Model result.
We include processes of coupling of photons to individual nucleons
(protons and neutrons) in addition to coherent coupling to the
whole nuclei (called standard approach here). 
Both elastic (nucleon in the ground state) and inelastic 
(nucleon in an excited state) in the couplings of photons to nucleons 
are taken into account. The inelastic nucleon fluxes are calculated
using CT18qed photon in nucleon PDFs. The inelastic photon fluxes are
shown and compared to standard photon fluxes in the nucleus.
In addition, we show the ratio of the inelastic corrections to 
the standard contribution as a function of diphoton invariant mass 
and photon rapidity difference. 
We find that for the ATLAS acceptance region the inelastic 
corrections grow with $M_{\gamma \gamma}$ and rapidity difference. 
Our results indicate that the inelastic contributions can 
be of the order of 20-40 \%  
of the traditional (no nuclear excitation) predictions. Uncertainties due to factorization scale choice are quantified.

\end{abstract}



\maketitle

\vspace{1cm}

\section{Introduction}
Over the last years, the study of photon - induced interactions in proton - proton, proton - nucleus and nucleus - nucleus collisions became a reality, allowing e.g. to observe for the first time the Light-by-Light (LbL) scattering as well as derive important constraints in several scenarios beyond the Standard Model (BSM) physics \cite{BSM}. The basic idea in the photon-induced processes is that an ultra-relativistic charged hadron (proton or nucleus)
 gives rise to strong electromagnetic fields, such that the photon stemming from the electromagnetic field
of one of the two colliding hadrons can interact with one photon of
the other hadron (photon-photon process) or can interact directly with the other hadron (photon-hadron
process) \cite{upc}. In particular, the cross - section for the LbL process in a hadronic collision can be expressed schematically as follows \cite{epa},
\begin{eqnarray}
\sigma^{LbL} (\sqrt{s_{NN}})   
&\propto&    f_{\gamma/h_1}(x_1) \otimes f_{\gamma/h_2}(x_2) \otimes  \, \hat{\sigma}\left[\gamma \gamma \rightarrow \gamma \gamma ; W_{\gamma \gamma} \right] 
  \,\,\, ,
\label{Eq:LbL}
\end{eqnarray}
where  $\sqrt{s_{NN}}$ is the center - of - mass energy of the $h_1h_2$ collision, $f_{\gamma/h_i}$ is the photon distribution function associated with the hadron $i$ and $x_i$ ($i = 1, \, 2$) are the fractions of the hadron energy carried by the photon. Moreover, $\hat{\sigma}$ represents the cross-section for the LbL scattering for a given photon - photon center - of - mass energy $W_{\gamma \gamma}$. From Eq.~(\ref{Eq:LbL}) one has that a basic ingredient of the calculation is the photon distribution function of the hadron.

For a charged pointlike fermion, the photon distribution function was calculated almost one hundred years ago by Fermi \cite{Fermi}, Williams \cite{Williams} and Weizs\"acker \cite{Weizsacker}. In contrast, the calculation of non-pointlike particles is still a subject of intense study. 
In recent years, the computation of higher - order QCD and electroweak (EW) corrections for hadronic processes have motivated huge progress in the determination of the photon distribution in the proton \cite{Gao:2017yyd,Amoroso:2022eow}, and several groups have derived such distribution, e.g. by solving the QED - corrected DGLAP equations \cite{lux1,lux2,Bertone:2017bme,Harland-Lang:2019pla,Cridge:2021pxm,Xie:2021equ,NNPDF:2024djq}.  In general, the photon content of the proton at a given scale $\mu$ is assumed to be expressed as a sum of two contributions:                        
\begin{eqnarray}
f_{\gamma/p} (x,\mu^2) = f_{\gamma/p}^{el} (x) + f_{\gamma/p}^{inel} (x,\mu^2) \,\,,
\label{total_proton}
\end{eqnarray}
where the elastic component, $f_{\gamma/p}^{el} (x)$, can be estimated analyzing the $p \rightarrow \gamma p$ transition taking into account the effects of the proton form factors, with the proton remaining intact in the final state.   On the other hand, the inelastic contribution, $f_{\gamma/p}^{inel} (x,\mu^2)$,  is associated to the transition $p \rightarrow \gamma X$, with $X \neq p$, and can be estimated taking into account the partonic structure of the proton, which can be a source of photons.  Currently, different groups have provided parametrizations for the photon distribution function,  which differ mainly in the methodology and data sets used to constrain this distribution \cite{lux1,lux2,Bertone:2017bme,Harland-Lang:2019pla,Cridge:2021pxm,Xie:2021equ,NNPDF:2024djq}.  In a similar way, the photon distribution of the neutron, 
\begin{eqnarray}
f_{\gamma/n} (x,\mu^2) = f_{\gamma/n}^{el} (x) + f_{\gamma/n}^{inel} (x,\mu^2) \,\,,
\label{total_neutron}
\end{eqnarray}
has also been derived in recent studies \cite{Cridge:2021pxm,Xie:2023qbn}. In principle, the contribution of the inelastic component can be suppressed, and the elastic one probed, in exclusive processes where events characterized by intact protons in the final state are tagged by forward detectors such as e.g. the AFP/ATLAS and CT-PPS/CMS detectors \cite{Adamczyk:2015cjy,Tasevsky:2015xya,CMS:2014sdw}.

For a nucleus,  $f_{\gamma/A}^{el} (x)$ is proportional to the squared charge of the ion ($Z^2$),  due to the coherent action of all protons in the nucleus. In contrast, $f_{\gamma/p}^{inel} (x)$ is proportional to the mass number $A$. As a consequence, for a heavy nucleus, the total photon distribution is expected to be dominated by its elastic component, which justifies that the analysis of photon - induced interactions in ultraperipheral heavy ion collisions have been estimated assuming that the photon distribution of the nucleus is given only by its elastic contribution, independently of the fact that the nuclei in the final state are not tagged by forward detectors. However, the dependence on the energy fraction of the elastic and inelastic components is distinct, with the inelastic one being dominant at large - $x$, which implies that such component can be important in some regions of the phase space.  Moreover, the measurements of the photon - induced interactions at the LHC is now entering the precision era, which means that subleading contributions can become important to describe the data.

\begin{figure}[t]
	\centering
	\begin{tabular}{ccccc}
\includegraphics[width=0.32\textwidth]{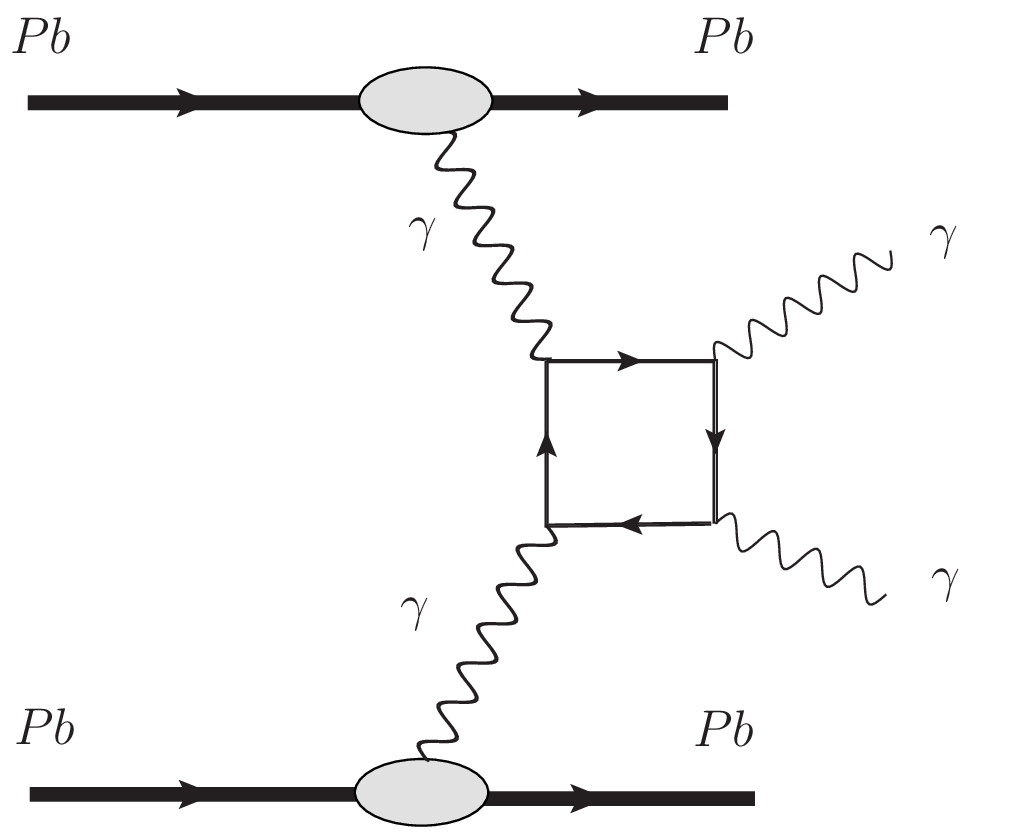} &
\includegraphics[width=0.32\textwidth]{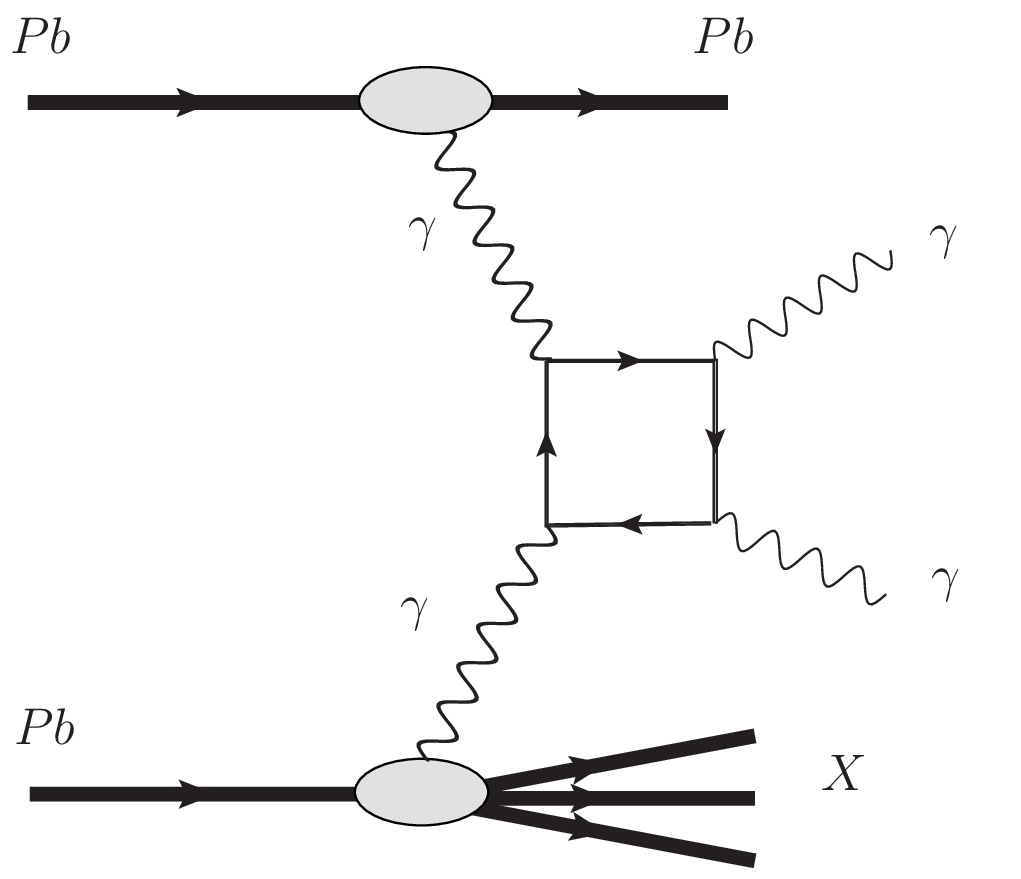} &
\includegraphics[width=0.32\textwidth]{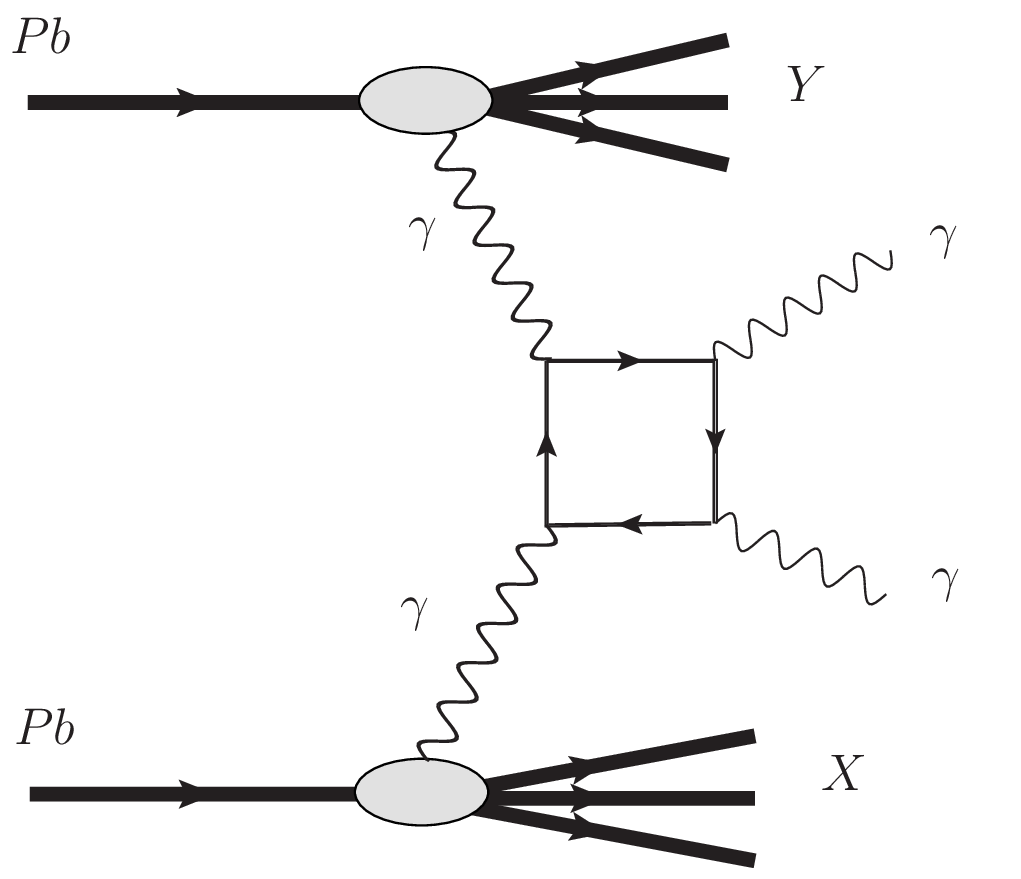} 
	\end{tabular}
\caption{Diagrams associated with the contribution of elastic (left panel), semi-elastic (central panel) and inelastic (right panel) processes for the LbL scattering in $PbPb$ collisions. }
\label{fig:diagram}
\end{figure}

The inclusion of the inelastic component in the calculation of LbL scattering in $PbPb$ collisions, implies that the cross-section will be given schematically by 
\begin{eqnarray}
\sigma^{LbL} (\sqrt{s_{NN}})   
&\propto&    f_{\gamma/Pb}^{el}(x_1) \otimes f_{\gamma/Pb}^{el}(x_2)  \otimes  \, \hat{\sigma}\left[\gamma \gamma \rightarrow \gamma \gamma ; W_{\gamma \gamma} \right]  \,\,  + \nonumber \\
& + & f_{\gamma/Pb}^{el}(x_1) \otimes f_{\gamma/Pb}^{inel}(x_2)  \otimes  \, \hat{\sigma}\left[\gamma \gamma \rightarrow \gamma \gamma ; W_{\gamma \gamma} \right]  \,\, + \nonumber \\
& + & f_{\gamma/Pb}^{inel}(x_1) \otimes f_{\gamma/Pb}^{el}(x_2)  \otimes  \, \hat{\sigma}\left[\gamma \gamma \rightarrow \gamma \gamma ; W_{\gamma \gamma} \right]  \,\, + \nonumber \\
& + & f_{\gamma/Pb}^{inel}(x_1) \otimes f_{\gamma/Pb}^{inel}(x_2)  \otimes  \, \hat{\sigma}\left[\gamma \gamma \rightarrow \gamma \gamma ; W_{\gamma \gamma} \right]   \,\,\, ,
\label{Eq:LbL_components}
\end{eqnarray}
which means that in addition to the contribution of elastic processes, represented in the left panel of Fig.  \ref{fig:diagram}, the total cross-section will receive contributions of semi-elastic (central panel) and inelastic (right panel) processes, where one or both of the incident nuclei will dissociate due to the photon emission, respectively. Our goal in this paper is to estimate, for the first time, the contribution of these two processes and compare our results with the experimental data. Our analysis is strongly motivated by the existing tension between the current data and theoretical predictions based on the elastic photon distribution, which is reduced but not eliminated when the next - to - leading order corrections to the LbL process are taken into account \cite{AH:2023ewe,AH:2023kor}.  As we will demonstrate below, our results indicate a sizable contribution of semi-elastic and inelastic processes.

This paper is organized as follows. In the next section, we briefly review the formalism needed to estimate the contribution of the elastic, semi-elastic and inelastic processes for the LbL scattering in $PbPb$ collisions. In section \ref{sec:results} we evaluate the cross-section including all contributions (elastic + semi-elastic + inelastic) and estimate the ratio between this result and the purely elastic one. Predictions for the dependence of the ratio on the diphoton invariant mass and rapidity difference will be presented, and a comparison of our predictions with the ATLAS experimental data will be performed. Finally, in section \ref{sec:conclusions} we summarize our main results and present conclusions.

\section{Formalism}
\label{sec:form}

In our calculations of the elastic process, we will follow closely the formalism detailed in Refs. \cite{Klusek-Gawenda:2016euz,Klusek-Gawenda:2019ijn, Jucha:2023hjg}, where the elementary cross
section, $\hat{\sigma}\left[\gamma \gamma \rightarrow \gamma \gamma ; W_{\gamma \gamma} \right]$, is calculated taken into account fermionic loops. For the  elastic photon flux associated with the nuclei, we  will  assume
\begin{eqnarray}
f_{\gamma/Pb}^{el} (x) & = & \frac{\alpha Z^2}{\pi x}\left\{2\xi K_0(\xi)K_1(\xi) - \xi^2[K_1^2(\xi) - K_0^2(\xi)] \right\}  \,\,,  
\end{eqnarray}
where $K_0$ and $K_1$ are the modified Bessel functions and $\xi = x M_A b_{min}$, with $b_{min} = R_A$. On the other hand, for the calculation of  semi-elastic and inelastic processes, we will assume that  inelastic photon flux of the nucleus is given by
\begin{eqnarray}
f_{\gamma/Pb}^{inel} (x,\mu^2) & = & Z \times f_{\gamma/p} (x,\mu^2) + (A - Z) \times f_{\gamma/n} (x,\mu^2) \,\,.
\end{eqnarray}
In other words, we will consider that the elastic component of the photon distribution is given by the Weiz\"acker - William distribution for a pointlike object \cite{upc} and the inelastic one is given by the incoherent sum of the photon distributions associated with the proton and neutron. It is important to emphasize that in our analysis, we will take into account both the elastic and inelastic contributions for the photon distributions of the nucleons. The distributions $f_{\gamma/p} (x,\mu^2)$ and $f_{\gamma/n} (x,\mu^2)$ will be estimated using the CT18qed parametrization \cite{Xie:2021equ,Xie:2023qbn}, assuming that the hard scale $\mu$ is the invariant mass of the produced diphoton system $M_{\gamma\gamma}$.

\begin{figure}[!b]
    \centering
    (a)\includegraphics[width=0.45\linewidth]{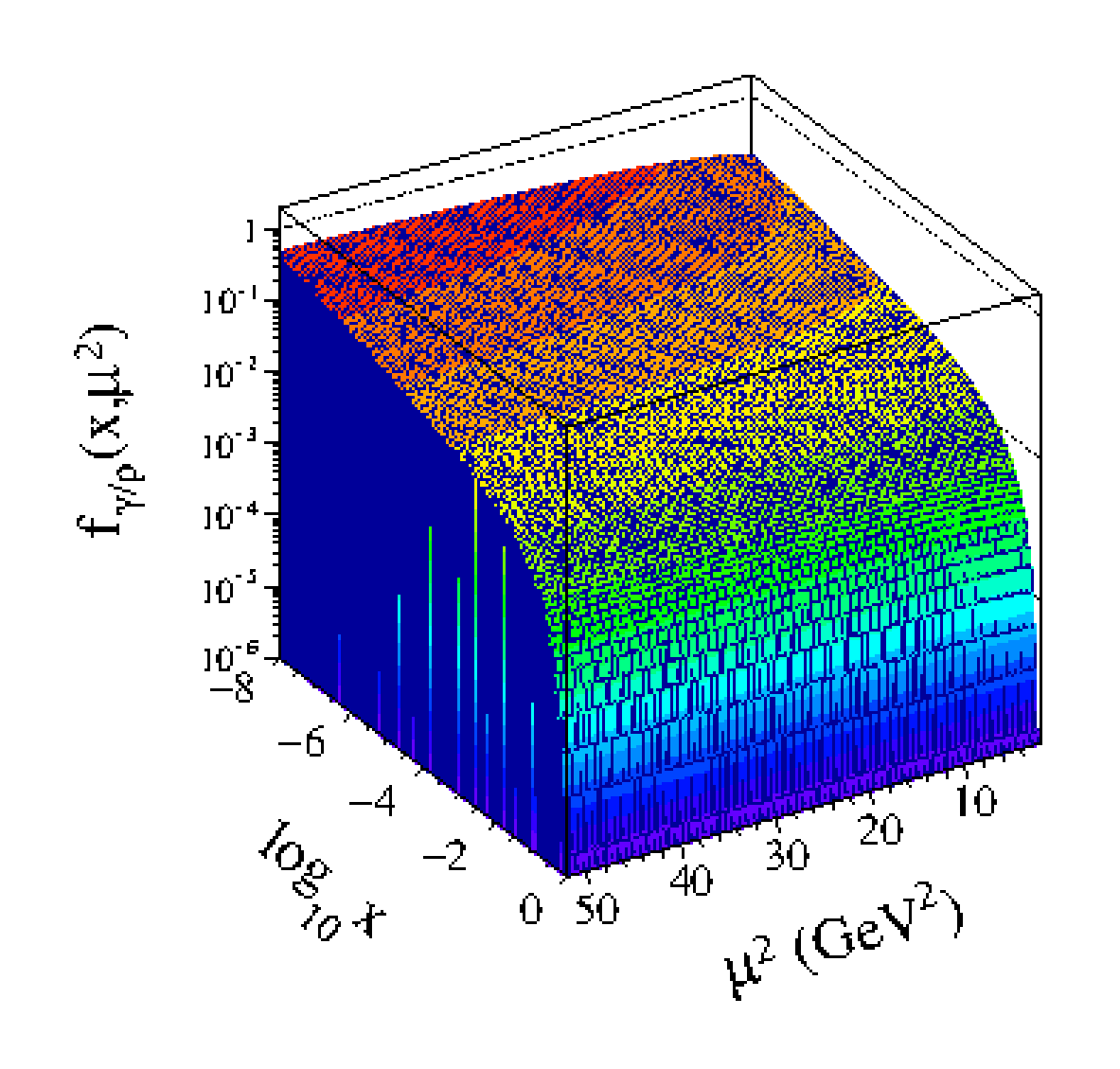}
    (b)\includegraphics[width=0.45\linewidth]{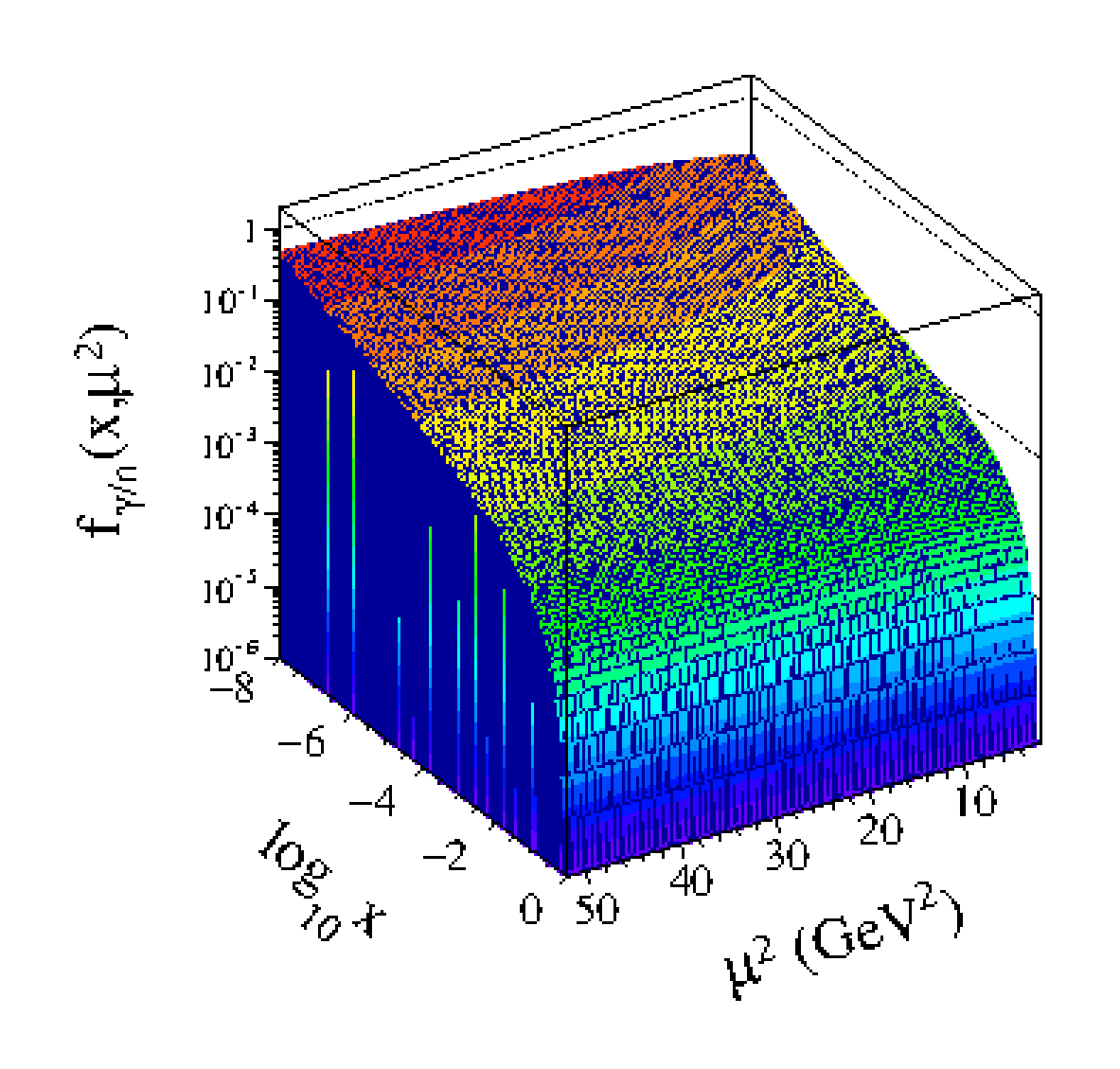}
    \caption{Photon distribution (sum of inelastic and elastic components) in (a) proton and (b) neutron as a function of $\log_{10}x$ and $\mu^2$.}
    \label{fig:f_photon_2dim}
\end{figure}
In Fig.~\ref{fig:f_photon_2dim} we show photon distribution in proton (a) and neutron (b) as a function of $\log_{10}x$ and the scale $\mu^2$. 
Such photon distributions have been obtained using the LUXqed method \cite{lux1, Manohar:2017eqh}, which computes the PDFs using only information from electron - proton scattering data, and taken into account the elastic (nucleon in the ground state) and inelastic (nucleon in an excited state) contributions. There is a fairly weak dependence on the scale. One has that the main difference between the photon PDFs for proton and neutron occurs at small hard scales, which is directly associated with the distinct contribution of the elastic photon flux, and large $x$ which is associated with different quark content of the proton and neutron. 

\begin{figure}[!h]
    \centering
    \includegraphics[width=0.45\linewidth]{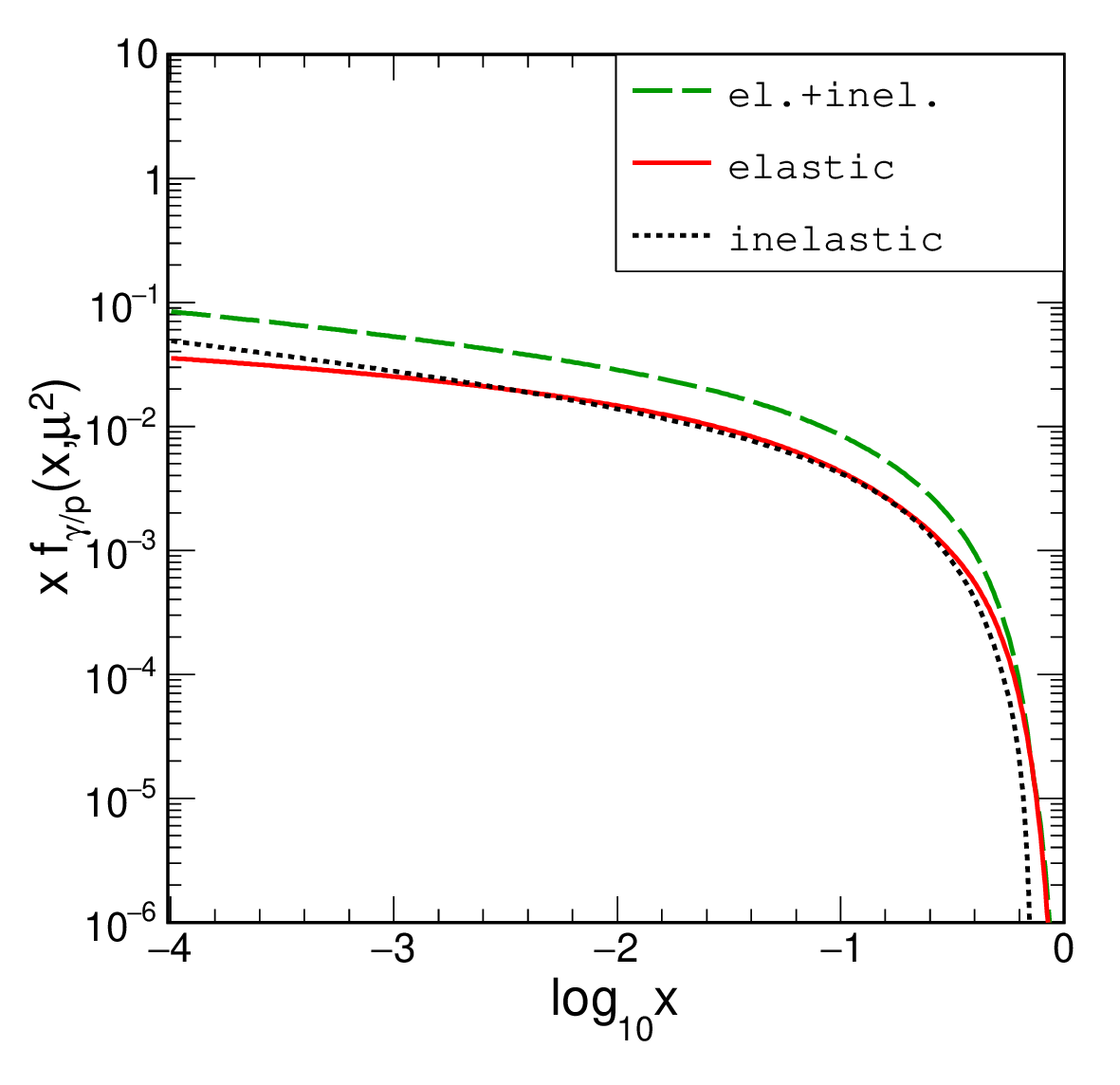}
    \includegraphics[width=0.45\linewidth]{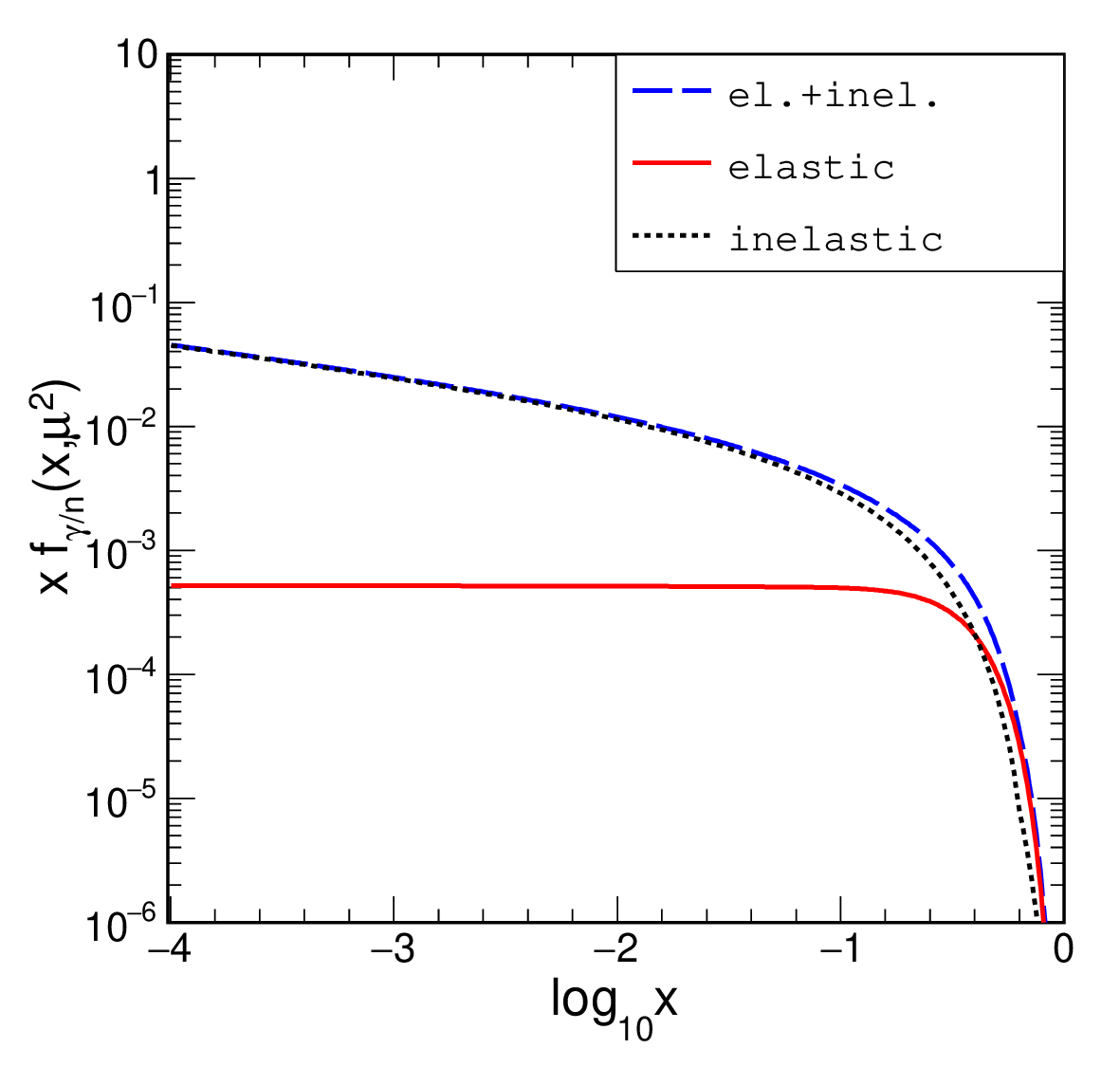}
    \caption{Elastic and inelastic photon distributions for (a) proton and (b) neutron as a function of $\log_{10}x$. In this plot, we assumed $\mu^2=25$~GeV$^2$.}
    \label{fig:xf_log_n_p}
\end{figure}

The nucleon (proton and neutron) flux is a sum of inelastic and elastic contribution. In the so-called LUXqed approach, not only inelastic but also elastic contribution is $\mu^2$-scale dependent \cite{Xie:2021equ,Xie:2023qbn}. 
The inelastic and elastic contributions are shown separately in Fig.~\ref{fig:xf_log_n_p}. Here we have taken $\mu=5$~GeV. The elastic component is expressed in terms of the electric and magnetic form factors, with the magnetic one being dominant in the neutron case due to the zero electric charge, in contrast to the proton case where both contributions are important. While for proton the elastic contribution is comparable to the inelastic one, for neutron the elastic contribution is rather small ($G_E \approx 0$). In contrast, the inelastic component is determined by the corresponding structure functions and has a similar evolution with $\mu^2$.

\begin{figure}
    \centering
    \includegraphics[width=0.45\linewidth]{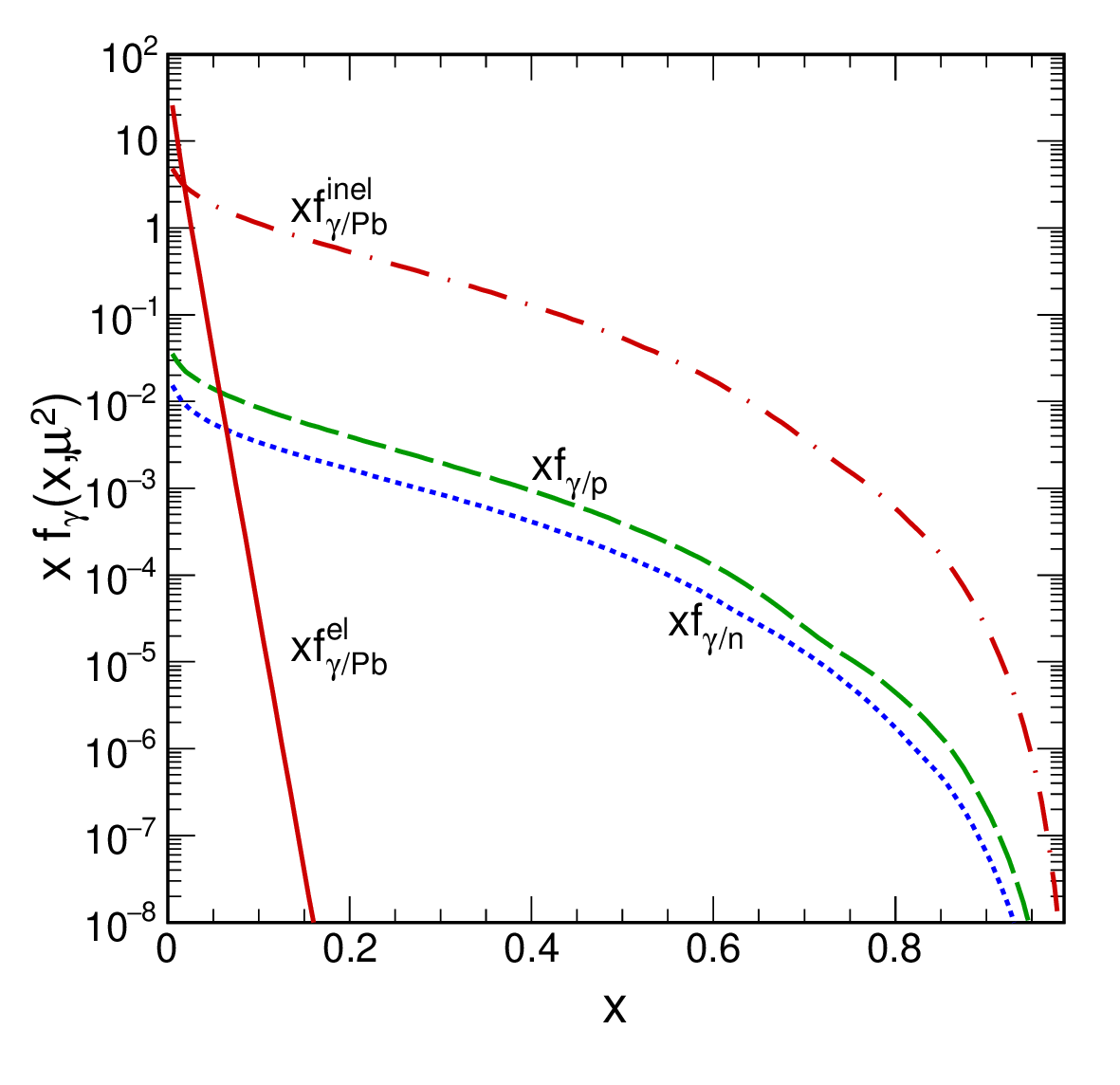}
    \caption{Elastic and inelastic photon distributions for the lead nucleus as a function of  $x$. For comparison, the proton and neutron distributions are also presented by the dashed and dotted line, respectively. In this plot, we assumed $\mu^2=25$~GeV$^2$.}
    \label{fig:xf}
\end{figure}

In Fig. \ref{fig:xf} we present a comparison  between the
elastic and inelastic photon fluxes of the nucleus. For completeness, the proton and neutron photon fluxes are also presented. While the  inelastic photon flux for a nucleus is proportional to $A$, the elastic photon spectrum is proportional to $Z^2$ and $1/x$, which implies that it dominates for a large nuclei and small values of $x$. However, as the inelastic flux is determined by the proton and neutron photon PDFs, its $x$ - behavior implies that it becomes dominant for large values of $x$. It is important to emphasize that the value of $x$ in which 
$f_{\gamma/Pb}^{el}  = f_{\gamma/Pb}^{inel}$ is dependent on the value of hard scale $\mu$. These results indicate that if the particle production by $\gamma \gamma$ interactions in UPHICs is dominated by the photons carrying a large value of $x$, we can expect a nonnegligible contribution of the semi-elastic and inelastic contributions. In the next Section, we will focus on the calculation of these contributions for the LbL scattering.

%

%

\section{Results for nuclear cross-sections}
\label{sec:results}

In what follows, we will estimate the LbL cross - section considering the sum of the elastic, semi-elastic and inelastic contributions, as described in Eq. (\ref{Eq:LbL_components}), and compare it with the results derived considering only the elastic term, as usually performed in the literature. We will consider $PbPb$ collisions at $\sqrt{s_{NN}} = 5.02$ TeV and  the kinematic cuts used by the ATLAS Collaboration, which assumed in its analysis that the photons in the final state have a transverse momentum $p_{t,1/2} >$ 2.5 GeV, and the protons are produced in the rapidity range -2.4 $< y_{1/2} <$ 2.4 and the diphoton invariant mass 
$M_{\gamma \gamma}$ is larger than 5 GeV.
In evaluation of the cross sections we take the scale parameter in the inelastic fluxes to be $\mu=M_{\gamma\gamma}$.
In a forthcoming study, we will investigate the impact of the semi-elastic and inelastic contributions considering different choices of cuts, which can be useful for the experimental collaborations to define the selection of events and quantify the magnitude of the SM background in the searching for new physics.

\begin{figure}[!h]
    \centering
(a)    \includegraphics[width=0.45\linewidth]{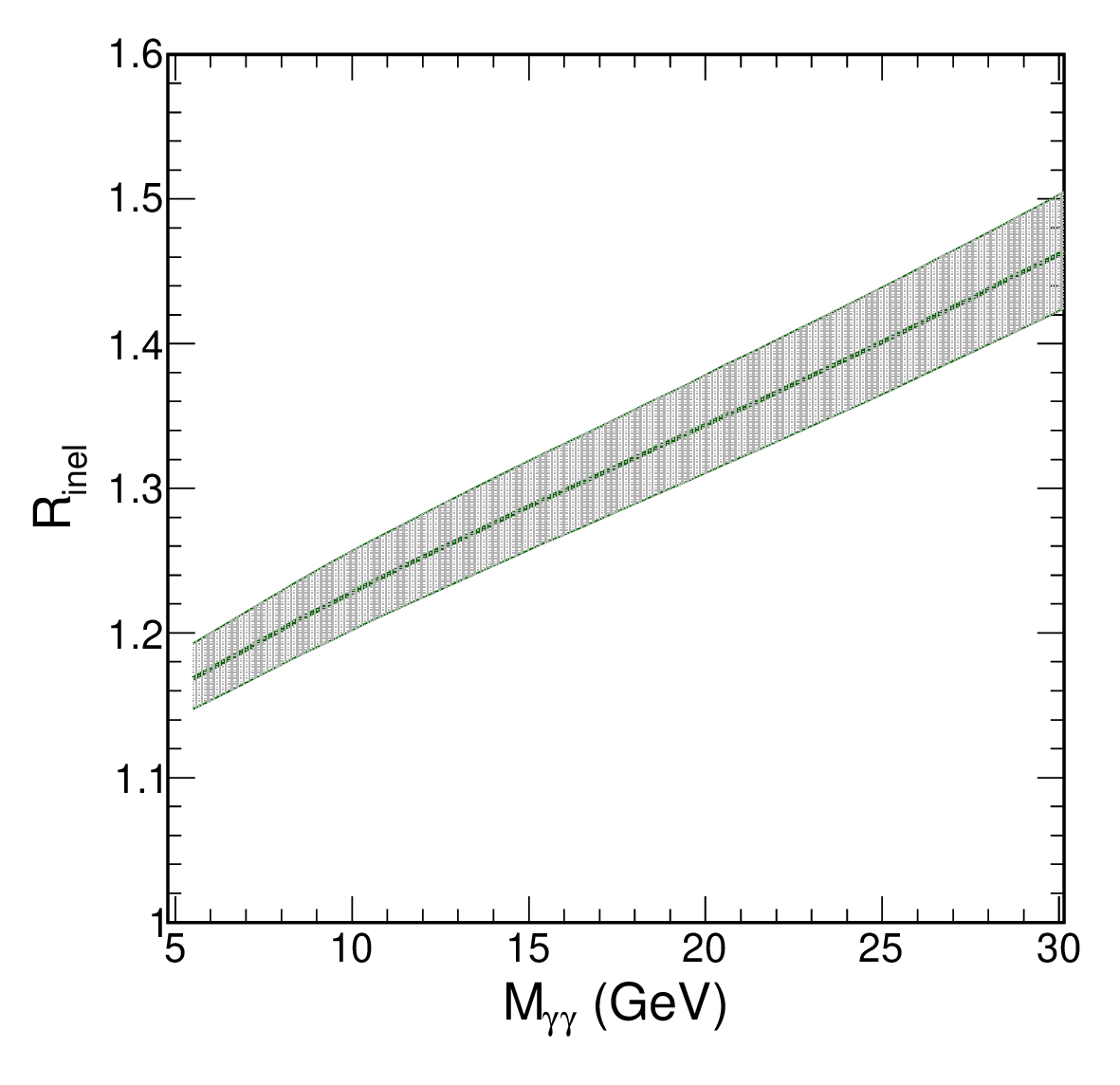}
(b)    \includegraphics[width=0.45\linewidth]{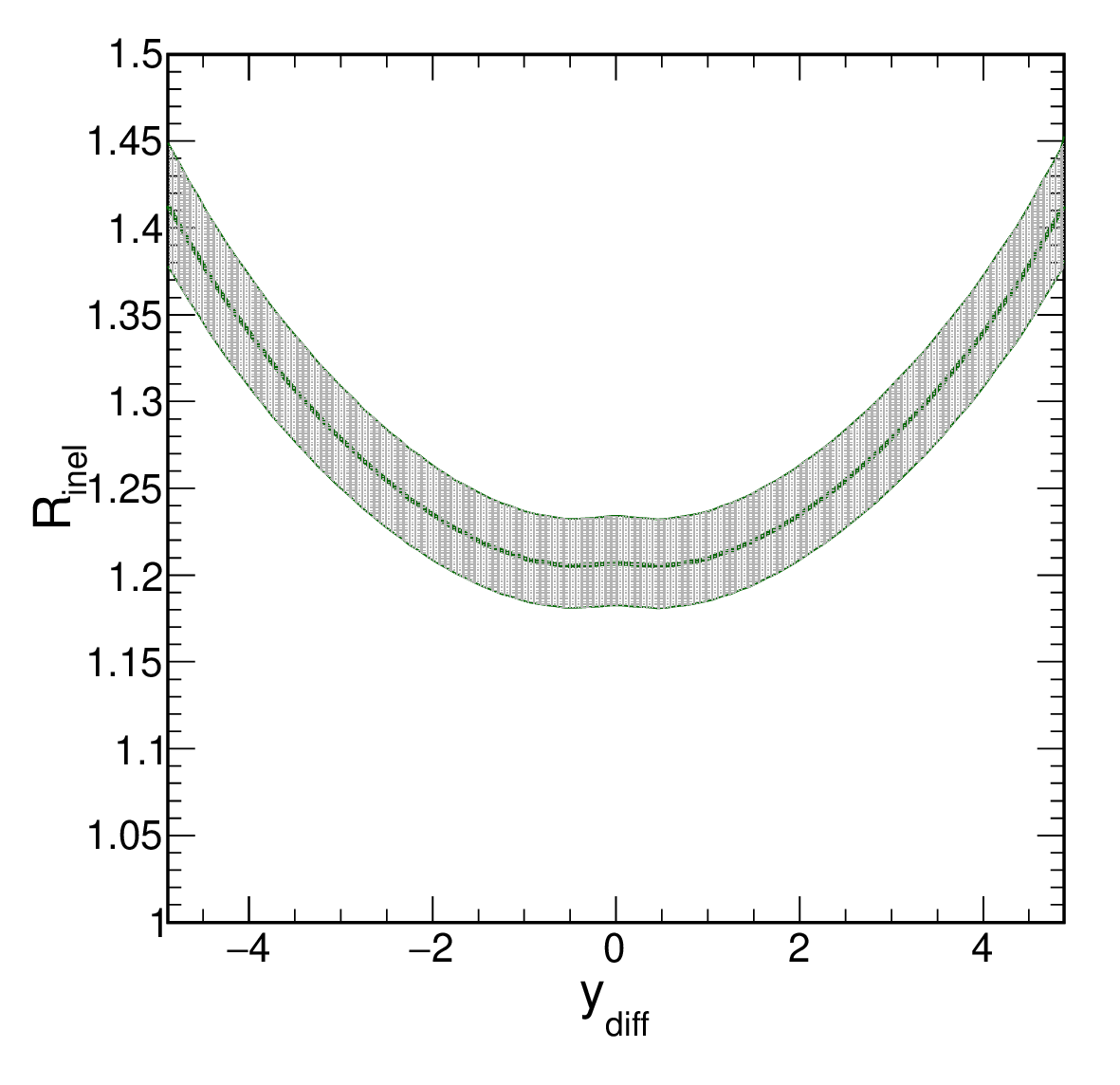}
    \caption{Dependence of the ratio $R_{inel}$, defined in
      Eq. (\ref{Eq:ratio}), on (a) the diphoton invariant mass and (b)
      rapidity difference. The uncertainty band due to the choice of factorization scale is shown
in addition.} 
    \label{fig:ratio_ela_ela}
\end{figure}

In order to quantify the impact of the semi-elastic and inelastic contributions, we will estimate the ratio defined by
\begin{eqnarray}
    R_{inel} \equiv \frac{d\sigma^{LbL}[\mbox{elastic + semi-elastic + inelastic}]}{d\sigma^{LbL}[\mbox{elastic}]} \,\,.
    \label{Eq:ratio}
\end{eqnarray}
In Fig. \ref{fig:ratio_ela_ela} we present the dependence of the ratio on the diphoton invariant mass (a) and rapidity difference $y_{diff} = y_1 - y_2$ 
(b), where $y_1$ and $y_2$ are rapidities of the first and 
second photon, respectively. Our results indicate that the ratio depends
on both $M_{\gamma \gamma}$ and $y_{diff}$. 
The results strongly suggest that the semi-elastic and inelastic 
contributions for the LbL scattering are sizable in the 
kinematical range probed by the ATLAS Collaboration.
We show in addition uncertainty band calculated by modifying
factorization scale between $M_{\gamma \gamma}^2/2$ and
$2 M_{\gamma \gamma}^2$.

\begin{figure}[!h]
    \centering
    \includegraphics[width=0.45\linewidth]{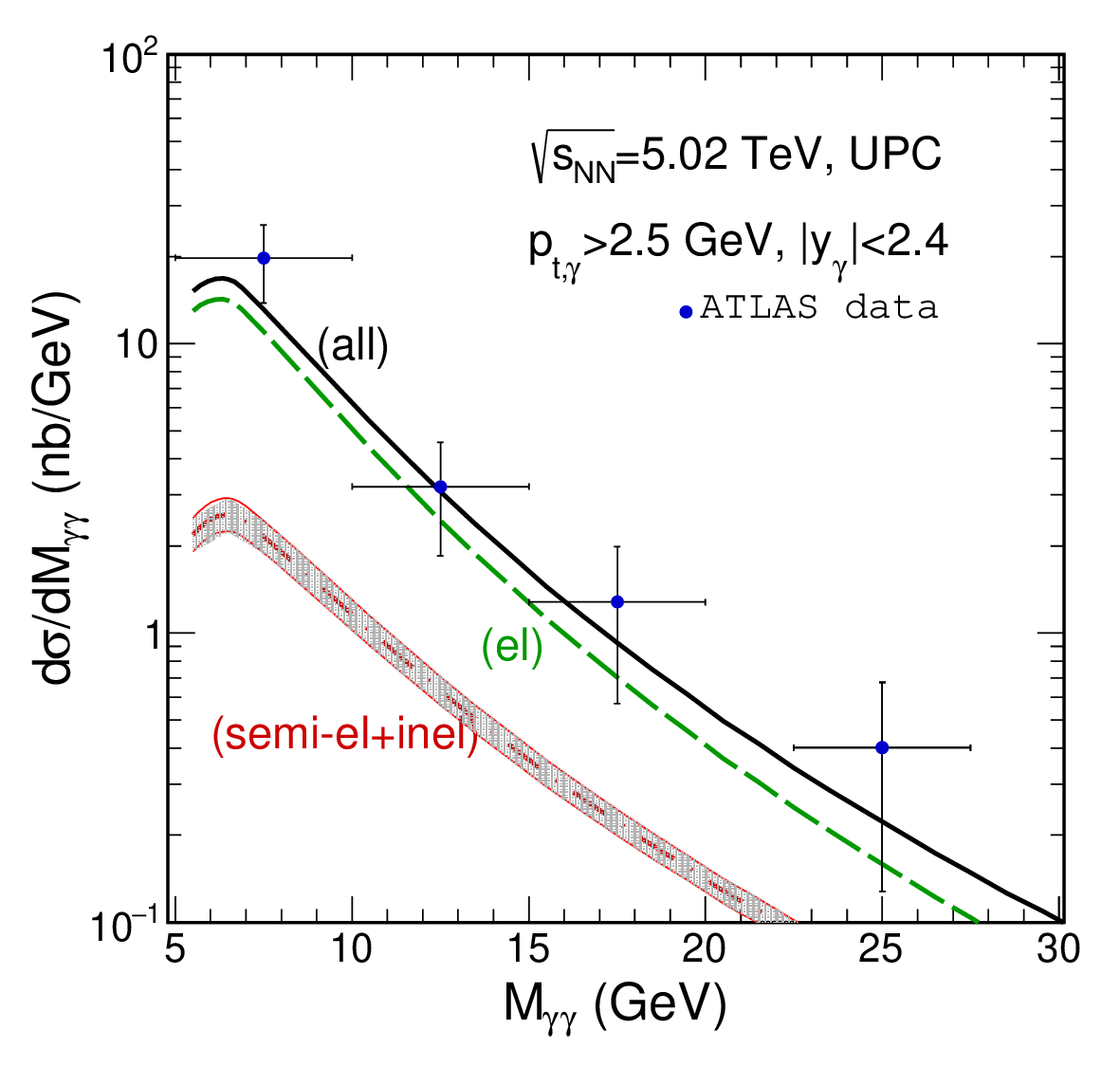}
    \caption{Differential cross-section for the LbL scattering in UPHICs
      as a function of diphoton invariant mass. Experimental ATLAS data
      are from Ref. \cite{ATLAS:2020hii}. 
We show uncertainty band due to the choice of factorization scale.}
    \label{fig:Comparison_data}
\end{figure}

In Fig. \ref{fig:Comparison_data} we compare our predictions for the LbL scattering in ultraperipheral $PbPb$ collisions with the ATLAS experimental data \cite{ATLAS:2020hii}. For comparison, we present separately the predictions associated with the elastic contribution (green dashed curve) and the sum of the semi-elastic and inelastic contributions (red dashed - dotted curve). As expected from the analysis of the ratio $ R_{inel} $, the new contributions are of the order of 20-40 \% of the elastic one at small values of $M_{\gamma \gamma}$. The sum of all contributions is also presented in Fig. \ref{fig:Comparison_data} (black solid line). One has that inclusion of the semi-elastic and inelastic contributions slightly improves the description of the ATLAS data.
The uncertainty band, $\mu^2 \in (M^2_{\gamma\gamma}, 2M^2_{\gamma\gamma})$ is shown for the inelastic contributions.

\section{Summary}
\label{sec:conclusions}

The current estimations of cross-section for the LbL scattering in ultraperipheral heavy ion collisions, $A A \to A A \gamma \gamma$, using the state-of-the-art nuclear
photon fluxes give predictions which are somewhat smaller than the measured ones by the ATLAS and CMS Collaborations,  although the current experimental statistics are not sufficient
for a definite conclusion.

In the present paper, we have considered new mechanisms in which
one or both initial photons couple rather to individual 
nucleons, protons or neutrons, instead of the coherent coupling 
to the nucleus as a whole. In our analysis,  we included both
$N \to N$ (elastic) and $N \to X$ (inelastic) nucleon contributions.
The nucleon inelastic contribution is calculated using a recent
CTEQ-TEA parametrization of the photon PDF in the proton and neutron.

We have estimated the cross-sections for the nuclear semi-elastic and 
inelastic contributions, and compared with the standard calculations 
of the elastic component.
We have found that the new contributions are of the order of 20-40 \%
compared to the elastic one, being dependent on the diphoton invariant
mass  $M_{\gamma \gamma}$ and rapidity difference $y_{diff} = y_1 -
y_2$. 
The found nuclear corrections give contributions to the measured 
cross-section that seem welcomed to understand the potentially missing 
strength with respect to the ATLAS or CMS data.
The inelastic corrections discussed here are much bigger than the NLO corrections 
to $\gamma \gamma \to \gamma \gamma$ \cite{AH:2023kor} and can be
difficult to eliminate experimentally. 

In general, the nuclear semi-elastic and inelastic contributions have a
unique impact on the final state due to emission 
of particles (protons, neutrons, pions, etc.) from an excited
nucleus. The analysis of the final state requires constructing a dedicated
Monte Carlo code that may be rather complicated, and its construction goes far 
beyond the present preliminary/exploratory analysis. 
However, it is important to emphasize that the contribution of 
semi-elastic processes for the particle production in $pp$ collisions
was implemented recently in the Superchic Monte Carlo generator
  \cite{Harland-Lang:2022jwn}, as well as the modeling of the contribution
  associated with the ion dissociation in UPCs
  \cite{Harland-Lang:2023ohq}. In principle, the effects studied in this
  paper could be implemented in this MC generator by including the
  recent developments.


\begin{acknowledgments}
V. P. G. would like to thank the members of the Institute of Nuclear Physics Polish Academy
of Sciences in Krak\'ow for their warm hospitality during the completion of this study. 
V.P.G. was partially supported by CNPq, FAPERGS and INCT-FNA (Process No. 464898/2014-5). M.KG prepared the results as part of the research project no. DEC2021/42/E/ST2/00350 funded by the National Science Center.

\end{acknowledgments}

\hspace{1.0cm}

\end{document}